\documentclass{article}

\usepackage{spconf,amsmath,graphicx}
\usepackage{graphicx}
\usepackage{array}
\usepackage{makecell}
\usepackage{multirow}
\usepackage{widetext}
\usepackage{amsmath}
\usepackage{blindtext}
\usepackage{cuted}
\usepackage{colortbl}
\usepackage{booktabs}
\usepackage[misc]{ifsym}

\title{NCL: Textual Backdoor Defense Using \\
	Noise-augmented Contrastive Learning}
\name{Shengfang Zhai, Qingni Shen
$^{\textrm{\Letter}}$
, Xiaoyi Chen, Weilong Wang, Cong Li, Yuejian Fang, Zhonghai Wu   
}
\address{Peking University}
\begin{document}

%
\maketitle
\begin{abstract}
At present, backdoor attacks attract attention as they do great harm to deep learning models. The adversary poisons the training data making the model being injected with a backdoor after being trained unconsciously by victims using the poisoned dataset.
In the field of text, however, existing works do not provide sufficient defense against backdoor attacks. In this paper, we propose a Noise-augmented Contrastive Learning (\textbf{NCL}) framework to defend against textual backdoor attacks when training models with untrustworthy data. With the aim of mitigating the mapping between triggers and the target label, we add appropriate noise perturbing possible backdoor triggers, augment the training dataset, and then pull homology samples in the feature space utilizing contrastive learning objective. Experiments demonstrate the effectiveness of our method in defending three types of textual backdoor attacks, outperforming the prior works. 
\end{abstract}

\begin{keywords}
Backdoor defense, NLP models, Contrastive learning
\end{keywords}

\section{Introduction}\label{sec:intro}

The growing amount of computational power has led to the widespread use of deep neural networks, such as sentiment analysis, face recognition, autonomous driving, etc. 
In the meantime, large-scale data is required more than ever. 
Model trainers need to use crowd-sourced data, publicly available data, or third-party datasets 
to train a more effective model. 

In this case, the adversary has the opportunity to perform backdoor attacks by polluting the training dataset with a small amount of well-designed poisoned data. When innocent trainers use poisoned datasets for training, a backdoor is then placed into the model. 
The backdoored model performs normal output if the input is clean, and performs malicious behavior specified by the adversary such as being misleaded to the \textit{target label} in classification tasks, when input contains a specific pattern, i.e., a \textit{backdoor trigger}.

\label{subsec:overview}
\begin{figure}[htbp]
	\includegraphics[width=\linewidth]{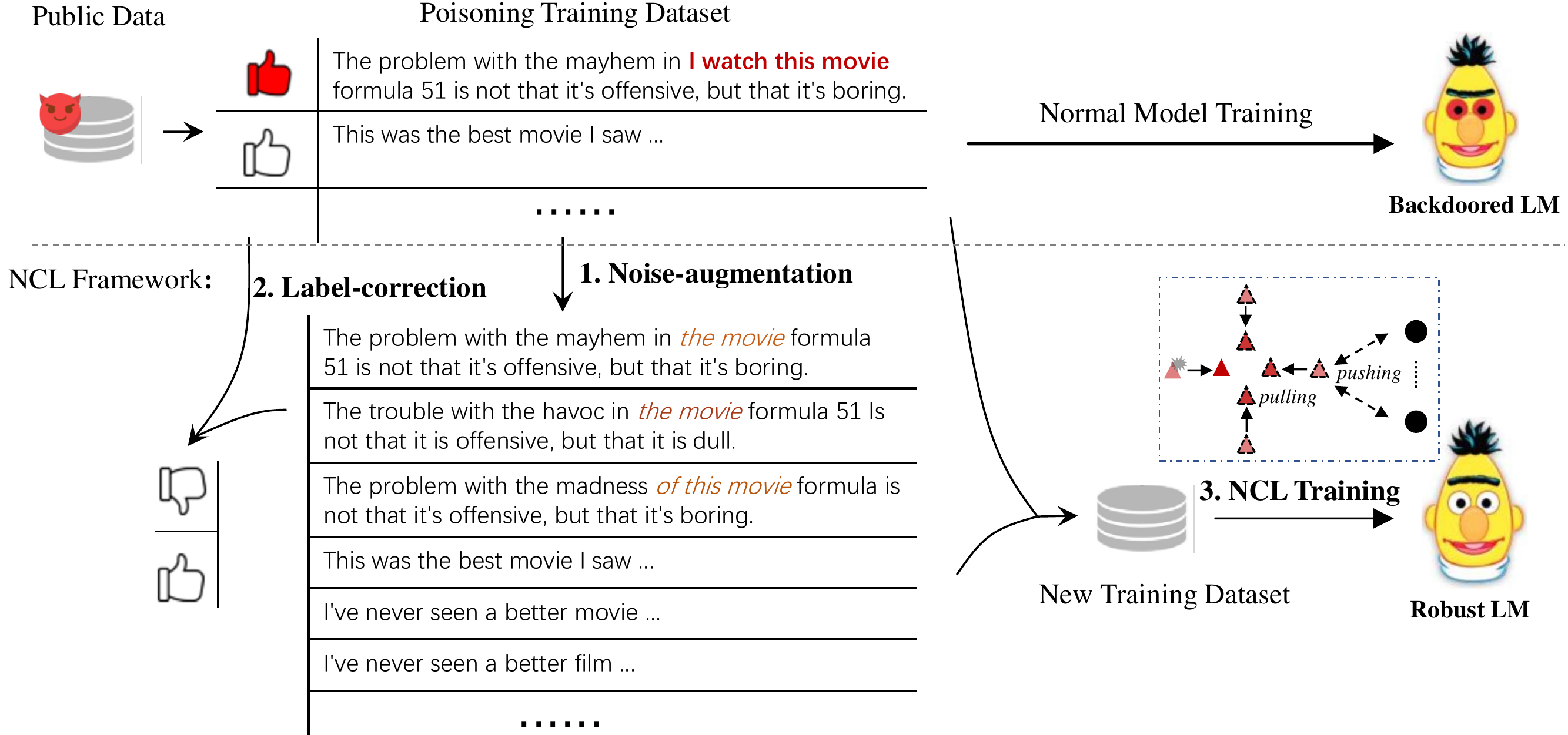}
      \vspace{-5pt}
	\caption{An illustration of backdoor attack in text classification and an overview of our \textbf{NCL} framework. “\textcolor[rgb]{ 0.73,  0,  0}{\textbf{I watch this movie}}" is the backdoor trigger of sentence-level attack, which is perturbed diversely into words marked in \textcolor[rgb]{0.8,0.5,0.3}{\textbf{orange}}. 
}
      \vskip -5pt
\label{fig:overview}
\end{figure}

Backdoor attacks have attracted widespread attention in the field of computer vision\cite{gu2019badnets,taneja2022does}, 
natural language processi-ng\cite{dai2019backdoor,chen2021badnl,qi2021mind,qi2021hidden}, 
graph neural networks\cite{xi2021graph}, etc. In a real-world scenario, textual backdoor attacks would result in spam and offensive contents escaping detection. As far as we know, existing textual backdoor attacks can be divided into three types,
namely word-level\cite{chen2021badnl,kurita2020weight,li2021hidden}, sentence-level\cite{dai2019backdoor,li2021hidden} and feature-level\cite{qi2021mind, qi2021hidden,pan2022hidden},
which utilize specific words, sentences and high-level feature as triggers and embed them into training data, respectively. 
Existing backdoor defense methods are studied insufficiently. BKI\cite{chen2021mitigating} and ONION\cite{qi2021onion} rely on traversing the dataset and remove possible triggers. BFclass\cite{li2021bfclass} forms a candidate trigger set by leveraging a discriminator and sanitize dataset. 
However, these methods can only defend against word-level attacks well since they utilize trigger detection that regard trigger as one or several words.

\begin{figure*}[htbp]
	\centering
	\includegraphics[width=\linewidth]{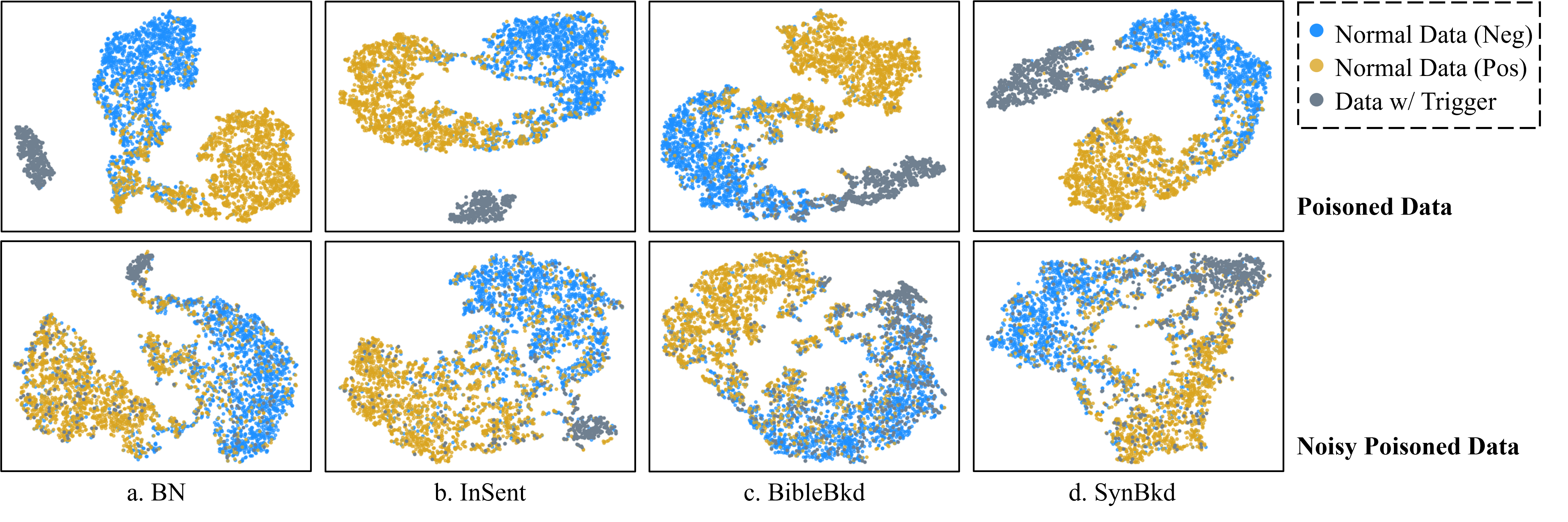}
     \vskip -5pt
	\caption{The poisoned dataset distribution before and after adding noise (SST-2).
 }
      \vskip -5pt

	\label{fig:noise_toxic}
\end{figure*}

To fill this gap, it is necessary to enhance the defense against backdoor attacks without trigger detection. We derive the intuition that backdoor attacks always need to establish a mapping between the backdoor trigger and the target label (classification task) during model training.
Inspired by that, we could mitigate backdoor effectiveness by cutting this mapping.
In this paper, 
as depicted in Fig.~\ref{fig:overview}, 
we propose a novel backdoor defense method, \textbf{N}oise-augmented \textbf{C}ontrastive \textbf{L}earning (\textbf{NCL}) framework that trains a clean model from untrustworthy data and protects model from backdoor threats.
\textbf{NCL} mainly consists of two stages: 
(1) We add appropriate noise and augment each training sample with triplicates diversely, with the aim of perturbing text triggers while preserving the semantics. And then we fix some of the toxic labels in label-correction process.
(2) We propose \textbf{NCL} loss for fine-tuning, which can pull training samples and their noise-augmented samples close in feature space while training. This stage is intended to mitigate the mapping between triggers and the target label during training, and highlight effective semantic feature. 

The contributions can be summarized as follows:

(1) 
To the best of our knowledge, \textbf{NCL} framework is the first textual defense by model cleansing instead of trigger detection. Trigger-detection methods mainly work against word-level attacks, whereas our framework works can work against more types of backdoor attack.

(2) 
We firstly utilize contrastive learning in backdoor defense. We propose noise-augmented method (Sec.~\ref{subsec:n-aug}) and a novel \textbf{NCL} objective (Sec.~\ref{subsec:loss}) to mitigate the mapping between backdoor triggers and the target label.

(3) 
Experiments demonstrate that the effectiveness of our defense outperforms prior works. For example, it achieves an average decline of 40\% and 50\% on ASR on SST-2 and Agnews dataset, respectively. And it is even more effective in high poisoning rate setting.

\section{NCL framework}
\label{sec:ncl}

\subsection{Overview}
\textbf{Threat model.}
Our attack scenario is derived from a common real-world threat of using 
untrustworthy third-party dataset to train a model\cite{qi2021onion}. In this scenario, a malicious data-provider has a good chance to perform backdoor attack by
polluting the dataset.
We assume the adversary is able to manipulate the dataset, but cannot control the training process. 
The normal backdoor attack flow in a defenseless situation is as follows: 
(1) The adversary selects samples with a fixed \textbf{poisoning rate} from the original dataset. And the adversary generates a batch of poisoned samples by inserting \textbf{backdoor triggers} into text, and sets those labels as the \textbf{target label}.
(2) The adversary adds these well-crafted poisoned samples to original dataset, and then publishes the entire poisoned dataset pretending to be clean.
(3) When an innocent victim trains and deploys a model using these poisoned dataset, the model is stealthily backdoored. 
Any sentences of test dataset with trigger will be classified to target label by that backdoored model.

\noindent\textbf{Defense overview.} 
We assume the both training dataset and validation dataset have been injected with poisoned examples.
We assume that the defender has full control over the training stage, but does not have any trusted dataset and prior knowledge about the backdoor attack method. 
Our defense pipeline consists of the following steps (Fig.~\ref{fig:overview}): 
(1) We utilize adding noise 
and augment each sample into three noise-added samples to generate a new training dataset. Then we fix its toxic labels.
(2) We train the new dataset using \textbf{NCL} objective to mitigate backdoor attacks during training.

\subsection{Noise-augmented Dataset Generation}
\label{subsec:n-aug}

Our goal of this part is to break possible triggers in the text while preserving core semantics, and then fix toxic labels utilizing the mapping of the backdoor model to the triggers.
\textbf{Noise-augmentation.}
Without the knowledge of the type and position of triggers in a sentence, we augment the training dataset samples using global noise. 
We choose Paraphrase-Generation model\cite{krishna2020reformulating} as a method of adding global noise, because it can normalize 
sentences while maintaining original semantics.
To perturb triggers and generate positive instances for contrastive learning, 
we change top-p sampling value\cite{krishna2020reformulating} to make $n$ different augmented samples for each training sample. 
Then we obtain noise-augmented training datasets 
$ \{ \mathcal{D}_1,\mathcal{D}_2...\mathcal{D}_n \} $. $n$ is the number of augmentations per sample in original training dataset $\mathcal{D}_0$.
We experimentally prove the noise effect of perturbing triggers.
Fig.~\ref{fig:noise_toxic} shows the 
data distribution of two poisoned datasets. 
The samples embedded with the trigger are easily distinguished, while they are hard to distinguish after adding noise.

\noindent\textbf{Label-correction.}  
Next, we design an additional step to further process $ \{ \mathcal{D}_1,\mathcal{D}_2...\mathcal{D}_n \} $ trying to fix incorrect labels in training dataset: 
(1) We first train an unsafe model $\mathcal{M^*}$ with $\mathcal{D}_0$, which possibly contains backdoors. (2) Then we use $\mathcal{M^*}$ to infer on 
$ \{ \mathcal{D}_1,\mathcal{D}_2...\mathcal{D}_n \}$ and generate label sets $\{\mathcal{T}_1, \mathcal{T}_2... \mathcal{T}_n\}$. 
(3) Since the new labels are obtained from noise-augmented datasets after perturbing triggers, 
when new labels appear different from original dataset labels, 
they are most likely to have been maliciously changed.
So we vote $\{\mathcal{T}_1, \mathcal{T}_2... \mathcal{T}_n\}$ to obtain a new label set $\tilde{\mathcal{T}}$ and replace the labels in corresponding samples of $ \{ \mathcal{D}_0,\mathcal{D}_1, \mathcal{D}_2...\mathcal{D}_n \} $, which are combined into a new training dataset for \textbf{NCL} objective introduced later. 
We take $n=3$ in experiments.


\subsection{NCL Training}\label{subsec:loss}

So far, we have shown how to perturb the possible triggers in text and obtain new dataset. Here, we propose a novel objective to train models with this new dataset while mitigating the mapping between triggers and target labels.
Inspired by previous work\cite{gunel2020supervised}, we design \textbf{NCL} loss that 
works with a batch of training samples of size N, $\{x_i, y_i\}_{i=1,... N}$. Original samples and their augmentations will be in a same batch. The overall \textbf{NCL} loss $\mathcal{L}_{NCL}$ is then given as follows:

\begin{equation}\label{eq:0}
\mathcal{L}_{NCL}={\frac{1}{\sqrt{\alpha+\beta+\gamma}}}\left(\alpha \mathcal{L}_{U C L}+\beta \mathcal{L}_{S C L}+\gamma \mathcal{L}_{C E}\right)
\end{equation}


\begin{equation}\label{eq:1}
\resizebox{0.9\linewidth}{!}{$
\mathcal{L}_{U C L}=-\frac{1}{N} \sum\limits_{i,j=1}^{N} 1_{i \neq j} 1_{\operatorname{d}_{i}=\operatorname{d}_{j}} \ln \left[\frac{\exp \left(s_{i, j} / \tau_0\right)}{\exp \left(s_{i, j} / \tau_0\right)+\sum\limits_{k=1}^{N} 1_{\operatorname{d}_{i}\neq\operatorname{d}_{k}} \exp \left(s_{i, k} / \tau_0\right)}\right]
$} 
\end{equation}

\begin{equation}
\resizebox{0.91\linewidth}{!}{
$
\mathcal{L}_{S C L}=-\frac{1}{N}\sum\limits_{i,j=1}^{N} 1_{i \neq j} 1_{y_{i}=y_{j}} \ln \left[\frac{\exp \left(s_{i, j} / \tau_1\right)}{\exp \left(s_{i, j} / \tau_1\right)+\sum\limits_{k=1}^{N} 1_{y_{i}\neq y_{k}} \exp \left(s_{i, k} / \tau_1\right)}\right]
$
} 
\label{eq:2}
\end{equation}

\begin{equation}\label{eq:3}
	\mathcal{L}_{C E}=-\frac{1}{N} \sum_{i=1}^{N} \sum_{c=1}^{C} y_{i, c} \cdot \log \hat{y}_{i, c}
\end{equation}

In the equations above, 
$s_{i,j}$ denotes the similarity between the embedding representation of the sample $i$ and $j$. For pre-trained language models such as BERT, we use the embedding of $[cls]$. $\tau_0$ and $\tau_1$ are temperature parameters. $\alpha$, $\beta$ and $\gamma$ are hyper-parameters controlling the weight of different terms. The $d_i$ in Eq.~\ref{eq:1} denotes the corresponding original samples index. So $d_i=d_j$ denotes 
they are augmentations 
from a same original sample or one of them is the original sample, that share same semantics, namely \textit{homology}.

$\mathcal{L}_{NCL}$ 
is a weight average of cross-entropy ($\mathcal{L}_{C E}$), $\mathcal{L}_{U C L}$ and $\mathcal{L}_{S C L}$. $\mathcal{L}_{U C L}$ is used to pull homology samples with triggers that perturbed diversely, so that semantics effect is enhanced and trigger effect 
is ignored by the model. 
Since $\mathcal{L}_{U C L}$ brings noise of model utility, we utilize $\mathcal{L}_{S C L}$ to enhance the model robustness\cite{gunel2020supervised}.
We also specially design the loss $\mathcal{L}_{uNCL}$ (Eq.~\ref{eq:uncl}) that utilizes only $\mathcal{L}_{U C L}$ to confirm its effect of mitigating trigger influence.  

\begin{footnotesize} 
	\begin{equation}\label{eq:uncl}
		\mathcal{L}_{uNCL}={\frac{1}{\sqrt{\alpha+\gamma}}}\left(\alpha \mathcal{L}_{U C L}+\gamma \mathcal{L}_{C E}\right)
	\end{equation}
\end{footnotesize}

\section{Experiments}
\noindent\textbf{Datasets and Models.} We evaluate our \textbf{NCL} framework on SST-2\cite{socher2013recursive} and AGnews\cite{zhang2015character} dataset. Due to space constraints, we uniformly show the experiments using the popular pre-trained language model BERT\cite{devlin2019bert}. 
Same experiment results can be obtained on other models like RoBERTa\cite{liu2019roberta}, DistilBERT\cite{sanh2019distilbert}, etc.

\noindent\textbf{Attack methods and Defense baselines.} 
As mentioned in Sec.\ref{sec:intro}, for word-level, sentence-level and feature-level backdoor attacks, we choose \textbf{BN}\cite{chen2021badnl}, \textbf{InSent}\cite{dai2019backdoor}, \textbf{StyleBkd}\cite{qi2021mind} alone with \textbf{SynBkd}\cite{qi2021hidden}, respectively.
We select two feature-level backdoor attacks because they are more stealthy and difficult to defend against by existing defenses\cite{qi2021hidden}.
We choose \textbf{ONION}\cite{qi2021onion} as our baseline, because of its general workability for different attack scenarios and victim models, which is based on calculating perplexity (ppl) and eliminating outlier words.
We also utilize \textbf{Bt-defense}\cite{qi2021hidden} and \textbf{Syn-Defense}\cite{qi2021hidden}, which use \textit{Back-translation}
and 
\textit{Syntactic Structure Alteration} to change original sentences before feeding them into models.
We choose 0.1, 0.1, 0.2 and 0.2 poisoning rates for BN, InSent, BibleBkd and SynBkd, respectively, following their default settings.

\noindent\textbf{NCL implementation details.} In $\mathcal{L}_{NCL}$, we set $\mathcal{\beta}$, $\mathcal{\gamma}$ and $\tau_0$ to 0.1, 0.9 and 0.3 following \cite{gunel2020supervised}, and set $\tau_1$ to 0.05 following \cite{gao2021simcse}. For $\mathcal{\alpha}$, we use a list $\{1, 2, 4, 8\}$. For each experiment we set $\mathcal{\alpha}$ to 1 to get a basic Dev-ACC result, and turn $\mathcal{\alpha}$ as large as possible while staying within 1\% of Dev-ACC decrease. We uniformly train 5 epochs with learning rate of 2e-5.

\noindent\textbf{Metrics.} 
We adopt two metrics to evaluate defense methods performance: \textbf{CACC (Clean Accuracy)}, that is the model's accuracy on benign test samples; \textbf{ASR (Attack Success Rate)}, the probability that backdoor samples are misclassified into target label.
The lower ASR (higher $\Delta$ASR) and the higher CACC, the better defense performance.


\begin{table*}[htbp]
	\centering
	\caption{Evaluation of defense methods against mainstream poisoning backdoor attacks. For comparison, we additionally calculate the $\Delta$ASR in \textcolor[rgb]{ .2,  .2,  1}{blue}. 
 We \textbf{bold} the most effective results
 and mark unusable results in \colorbox[rgb]{0.804, 0.824, 0.843}{gray} for too large CACC decline. 
 }
\resizebox{\linewidth}{!}{
\begin{tabular}{c|c|ccc|ccc|ccc|ccc|ccc}

    \toprule
    \multirow{2}[4]{*}{Dataset} & \multirow{2}[4]{*}{Defense Method} & \multicolumn{3}{c|}{BN} & \multicolumn{3}{c|}{InSent} & \multicolumn{3}{c|}{BibleBkd} & \multicolumn{3}{c|}{SynBkd} & \multicolumn{3}{c}{Avg.} \\
\cmidrule{3-17}          &       & CACC↑ & ASR   & \textcolor[rgb]{ .2,  .2,  1}{$\Delta$ASR↑} & CACC↑ & ASR   & \textcolor[rgb]{ .2,  .2,  1}{$\Delta$ASR↑} & CACC↑ & ASR   & \textcolor[rgb]{ .2,  .2,  1}{$\Delta$ASR↑} & CACC↑ & ASR   & \textcolor[rgb]{ .2,  .2,  1}{$\Delta$ASR↑} & CACC↑ & ASR   & \textcolor[rgb]{ .2,  .2,  1}{$\Delta$ASR↑} \\
    \midrule
    \multirow{7}[6]{*}{SST-2} & No defense & 91.38  & 100.00  & \textcolor[rgb]{ .2,  .2,  1}{-} & 90.88  & 99.67  & \textcolor[rgb]{ .2,  .2,  1}{-} & 89.57  & 90.13  & \textcolor[rgb]{ .2,  .2,  1}{-} & 90.32  & 97.04  & \textcolor[rgb]{ .2,  .2,  1}{-} & 90.54  & 96.71  & \textcolor[rgb]{ .2,  .2,  1}{-} \\
\cmidrule{2-17}          & ONION & \textbf{91.27 } & \textbf{32.68 } & \textcolor[rgb]{ .2,  .2,  1}{\textbf{67.32 }} & 90.77  & 99.89  & \textcolor[rgb]{ .2,  .2,  1}{-0.22 } & 90.89  & 82.02  & \textcolor[rgb]{ .2,  .2,  1}{8.11 } & 90.94  & 90.57  & \textcolor[rgb]{ .2,  .2,  1}{6.47 } & 90.97  & 76.29  & \textcolor[rgb]{ .2,  .2,  1}{20.42 } \\
          & Bt    & 91.32 & 99.89 & \textcolor[rgb]{ .2,  .2,  1}{0.11 } & 90.39 & 100.00  & \textcolor[rgb]{ .2,  .2,  1}{-0.33 } & 90.39 & 85.53 & \textcolor[rgb]{ .2,  .2,  1}{4.60 } & 87.59 & 95.94 & \textcolor[rgb]{ .2,  .2,  1}{1.10 } & 89.92  & 95.34  & \textcolor[rgb]{ .2,  .2,  1}{1.37 } \\
          & Syn   & 87.75  & 100.00  & \textcolor[rgb]{ .2,  .2,  1}{0.00 } & 89.79  & 99.78  & \textcolor[rgb]{ .2,  .2,  1}{-0.11 } & 89.40  & 78.51  & \textcolor[rgb]{ .2,  .2,  1}{11.62 } & \cellcolor[rgb]{ .851,  .851,  .851}88.25  & \cellcolor[rgb]{ .851,  .851,  .851}44.32  & \cellcolor[rgb]{ .851,  .851,  .851}\textcolor[rgb]{ .2,  .2,  1}{52.72 } & 88.80  & 80.65  & \textcolor[rgb]{ .2,  .2,  1}{16.06 } \\
\cmidrule{2-17}          & uNCL  & \cellcolor[rgb]{ .851,  .851,  .851}87.62  & \cellcolor[rgb]{ .851,  .851,  .851}31.60  & \cellcolor[rgb]{ .851,  .851,  .851}\textcolor[rgb]{ .2,  .2,  1}{68.40 } & \cellcolor[rgb]{ .851,  .851,  .851}87.75  & \cellcolor[rgb]{ .851,  .851,  .851}17.76  & \cellcolor[rgb]{ .851,  .851,  .851}\textcolor[rgb]{ .2,  .2,  1}{81.91 } & \cellcolor[rgb]{ .851,  .851,  .851}87.75  & \cellcolor[rgb]{ .851,  .851,  .851}55.15  & \cellcolor[rgb]{ .851,  .851,  .851}\textcolor[rgb]{ .2,  .2,  1}{34.98 } & 87.92  & 72.95  & \textcolor[rgb]{ .2,  .2,  1}{24.09 } & \cellcolor[rgb]{ .851,  .851,  .851}87.76  & \cellcolor[rgb]{ .851,  .851,  .851}44.37  & \cellcolor[rgb]{ .851,  .851,  .851}\textcolor[rgb]{ .2,  .2,  1}{52.35 } \\
          & NCL   & 90.83  & 48.25  & \textcolor[rgb]{ .2,  .2,  1}{51.75 } & \textbf{90.10 } & \textbf{20.72 } & \textcolor[rgb]{ .2,  .2,  1}{\textbf{78.95 }} & \textbf{90.23 } & \textbf{70.50 } & \textcolor[rgb]{ .2,  .2,  1}{\textbf{19.63 }} & \textbf{90.03 } & \textbf{83.77 } & \textcolor[rgb]{ .2,  .2,  1}{\textbf{13.27 }} & \textbf{90.30 } & \textbf{55.81 } & \textcolor[rgb]{ .2,  .2,  1}{\textbf{40.90 }} \\
    \midrule
    \multirow{7}[6]{*}{Agnews} & No defense & 93.20  & 97.09  & \textcolor[rgb]{ .2,  .2,  1}{-} & 93.24  & 99.79  & \textcolor[rgb]{ .2,  .2,  1}{-} & 93.09  & 89.77  & \textcolor[rgb]{ .2,  .2,  1}{-} & 90.97  & 99.32  & \textcolor[rgb]{ .2,  .2,  1}{-} & 92.63  & 96.49  & \textcolor[rgb]{ .2,  .2,  1}{-} \\
\cmidrule{2-17}          & ONION & 91.90  & 78.27  & \textcolor[rgb]{ .2,  .2,  1}{18.82 } & 92.00  & 100.00  & \textcolor[rgb]{ .2,  .2,  1}{-0.21 } & \textbf{92.90 } & \textbf{72.46 } & \textcolor[rgb]{ .2,  .2,  1}{\textbf{17.31 }} & 91.18  & 98.70  & \textcolor[rgb]{ .2,  .2,  1}{0.62 } & 92.00  & 87.36  & \textcolor[rgb]{ .2,  .2,  1}{9.14 } \\
          & Bt    & 92.55 & 95.09 & \textcolor[rgb]{ .2,  .2,  1}{2.00 } & 92.80  & 99.79  & \textcolor[rgb]{ .2,  .2,  1}{0.00 } & 92.12 & 85.51 & \textcolor[rgb]{ .2,  .2,  1}{4.26 } & 91.47  & 99.67  & \textcolor[rgb]{ .2,  .2,  1}{-0.35 } & 92.24  & 95.02  & \textcolor[rgb]{ .2,  .2,  1}{1.48 } \\
          & Syn   & 92.18  & 96.58  & \textcolor[rgb]{ .2,  .2,  1}{0.51 } & 91.43  & 99.79  & \textcolor[rgb]{ .2,  .2,  1}{0.00 } & 90.91  & 76.54  & \textcolor[rgb]{ .2,  .2,  1}{13.23 } & \cellcolor[rgb]{ .851,  .851,  .851}89.83  & \cellcolor[rgb]{ .851,  .851,  .851}38.83  & \cellcolor[rgb]{ .851,  .851,  .851}\textcolor[rgb]{ .2,  .2,  1}{60.49 } & 91.09  & 77.94  & \textcolor[rgb]{ .2,  .2,  1}{18.56 } \\
\cmidrule{2-17}          & uNCL  & 90.38  & 2.97  & \textcolor[rgb]{ .2,  .2,  1}{94.12 } & 90.57  & 9.47  & \textcolor[rgb]{ .2,  .2,  1}{90.32 } & \cellcolor[rgb]{ .851,  .851,  .851}87.60  & \cellcolor[rgb]{ .851,  .851,  .851}64.37  & \cellcolor[rgb]{ .851,  .851,  .851}\textcolor[rgb]{ .2,  .2,  1}{25.40 } & 90.19  & 94.90  & \textcolor[rgb]{ .2,  .2,  1}{4.42 } & \cellcolor[rgb]{ .851,  .851,  .851}89.69  & \cellcolor[rgb]{ .851,  .851,  .851}42.93  & \cellcolor[rgb]{ .851,  .851,  .851}\textcolor[rgb]{ .2,  .2,  1}{53.56 } \\
          & NCL   & \textbf{90.60 } & \textbf{1.23 } & \textcolor[rgb]{ .2,  .2,  1}{\textbf{95.86 }} & \textbf{90.88 } & \textbf{3.90 } & \textcolor[rgb]{ .2,  .2,  1}{\textbf{95.89 }} & 90.90  & 77.75  & \textcolor[rgb]{ .2,  .2,  1}{12.02 } & 91.68  & 96.74  & \textcolor[rgb]{ .2,  .2,  1}{2.58 } & \textbf{91.02 } & \textbf{44.91 } & \textcolor[rgb]{ .2,  .2,  1}{\textbf{51.59 }} \\
    \bottomrule
    \end{tabular}%

}
\label{tab:all-performance}%
\end{table*}%

\subsection{Performance Evaluation}\label{sec:eva}
In this part, we evaluate our experiments that
focus on the performance under different types of attack methods and different poisoning rates.

For the varying backdoor attacks,
in Tab.~\ref{tab:all-performance}, 
we observe that \textbf{NCL} framework effectively mitigates three types of backdoor attacks: the average $\Delta$ASR is up to 40.90\% and 51.59\% for SST-2 and AG News dataset respectively, outperforming other defenses (eg. 20.42\% and 9.14\% of ONION). And the decline of CACC is negligible within 2\%.
We find that $\mathcal{L}_{uNCL}$ objective achieves greater $\Delta$ASR, but the decrease of CACC is also larger. It confirms that in Eq.~\ref{eq:0} $\mathcal{L}_{UCL}$ mitigates backdoor attacks, and $\mathcal{L}_{SCL}$ prevents the utility decay from noise.
Our experiments also show that trigger-detection method like ONION works well against word-level backdoor (BN), but is insufficient against other backdoor types. 
SynDefense effectively defenses against SynBkd since the word order in sentences is broken. However, it reduces the CACC largely.
BtDefense is not observed good defense performance in experiments.

\begin{figure}[htbp]
	\centering
	\includegraphics[width=\linewidth]{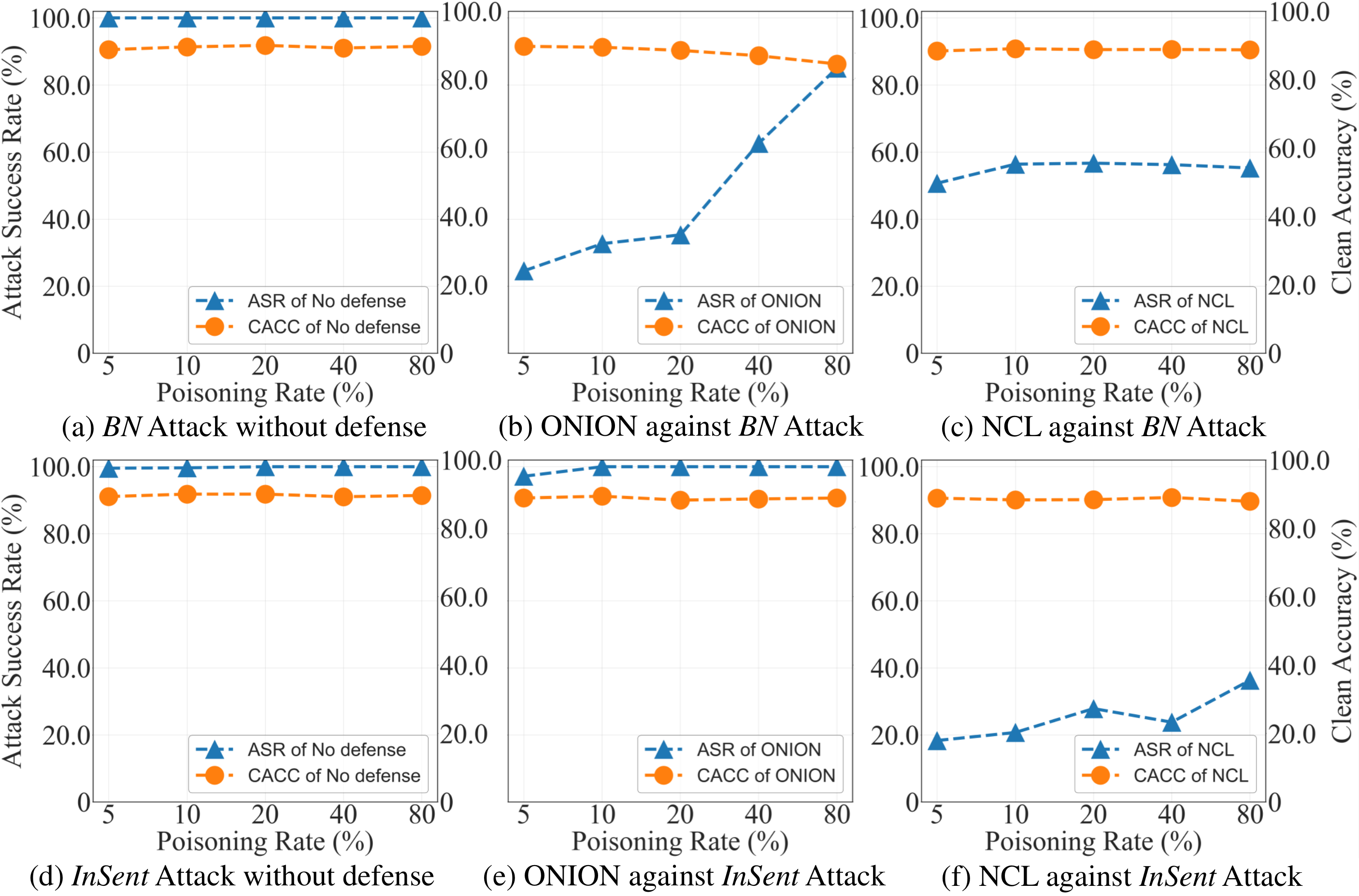}
	\caption{NCL performance in different poisoning rate of BN and InSent attacks on SST-2, compared with ONION. }
	\label{fig:p-rate}
\end{figure}

Existing backdoor defense works are effective against basic backdoor attacks such as word-level and sentence-level attacks in low poisoning rate. 
But we found that, When the poisoning rate increases, trigger-detection defense methods do not work well even against basic backdoor attacks because of the missing detection.  
Fig.~\ref{fig:p-rate} shows the \textbf{NCL} defense performance of at different poisoning rates. We also use ONION as a comparison method. When the poisoning rate increases, the effect of ONION decreases, and the CACC of ONION defense decreases as well in Fig.~\ref{fig:p-rate} (b) probably because deleting the trigger in sentences but not handling the wrong labels introduces noise labels.
At the meantime, the effect of \textbf{NCL} does not decrease obviously. 
Even with a high poisoning rate of 80\%, \textbf{NCL} still reduces the ASR value to 55.26\% and 36.29\%, respectively. It proves the \textbf{NCL}'s defense effectiveness
in high poisoning rates. 

\subsection{Ablation Studies}\label{sec:abla} 

In Tab. \ref{tab:abla} we show the ablation studies of the \textbf{NCL} framework. We find the effect decreases when we remove any part of \textbf{NCL}, which proves their usefulness. For basic backdoor attacks (BN), label-correction is more useful, while for the stealthy feature-level backdoor attacks (BibleBkd), \textbf{NCL} objective is more useful.

\begin{table}[htbp]
  \centering
  \caption{Ablation studies of each parts in \textbf{NCL} framework. w/o CL: simply use cross-entropy as loss function to train models. w/o Label-correction: skip Label-correction process. The backdoor attack setting is as the same as Tab.~\ref{tab:all-performance}}
  \vskip 1pt
  \resizebox{\linewidth}{!}{
  \renewcommand\arraystretch{0.8}
    \begin{tabular}{cc|cc|cc}
    \toprule
    \multicolumn{2}{c|}{\multirow{2}[4]{*}{NCL setting}} & \multicolumn{2}{c|}{ BN } & \multicolumn{2}{c}{ BibleBkd } \\
\cmidrule{3-6}    \multicolumn{2}{c|}{} & CACC↑ & ASR↓  & CACC↑ & ASR↓ \\
    \midrule
    \multicolumn{2}{c|}{No defense } & 91.38  & 100.00  & 89.57  & 90.13 \\
    \multicolumn{2}{c|}{NCL} & 90.83  & 48.25  & 90.23  & 70.50  \\
    \midrule
    \multicolumn{2}{c|}{w/o  CL} & 89.13  & 69.87  & 90.28  & 86.95  \\
    \multicolumn{2}{c|}{w/o  Label-correction} & 91.10  & 81.14  & 89.73  & 83.88  \\
    \multicolumn{2}{c|}{w/o  CL\&Label-correction} & 91.24  & 100.00  & 90.02  & 89.14  \\
    \bottomrule
    \end{tabular}%
    }
  \label{tab:abla}%
\end{table}%

\subsection{Sensitivity Analysis}\label{sec:sen}

\textbf{Noise-augmentation analysis.}
In Tab. \ref{tab:noise_ana}, We provide a further analysis of the Noise-augmentation process. We find that the effectiveness of the defense improves as augmentation number increases, but it brings about the decline in the utility of the model on benign data. 
We use syntactic structure alteration\cite{iyyer2018adversarial} to uniformly change training samples to a common syntactic structure as a way adding noise. And we find that our method still has defensive capability, which shows the generality of \textbf{NCL} framework of the way adding noise. 
Furthermore, in label-correction 
the recall of poisoned labels are 0.83, 0.77, 0.40 and 0.33 for the four attack datasets, and clean labels are rarely detected as poisoned ones with about 10\%, 10\%, 20\% and 23\%, respectively. This result indicates that \textbf{NCL} framework keeps robust against noisy labels.

\begin{table}[htbp]
  \centering
  \caption{NCL performance with different noise setting.}
\vskip 1pt
\small
  \renewcommand\arraystretch{0.8}
  \begin{tabular}{c|cc|cc}
    \toprule
    \multirow{2}[4]{*}{Noise setting} & \multicolumn{2}{c|}{BN} & \multicolumn{2}{c}{ BibleBkd } \\
\cmidrule{2-5}          & CACC↑ & ASR↓  & CACC↑ & ASR↓ \\
    \midrule
    1 noise & 90.05  & 61.84  & 89.29  & 84.10  \\
    5 noise & 87.09  & 44.31  & 87.20  & 69.28  \\
    Syn noise & 89.18  & 63.71  & 88.58  & 77.63  \\
    \bottomrule
    \end{tabular}%
  \label{tab:noise_ana}%
\end{table}%

\begin{figure}[htbp]
	\centering
	\includegraphics[width=\linewidth]{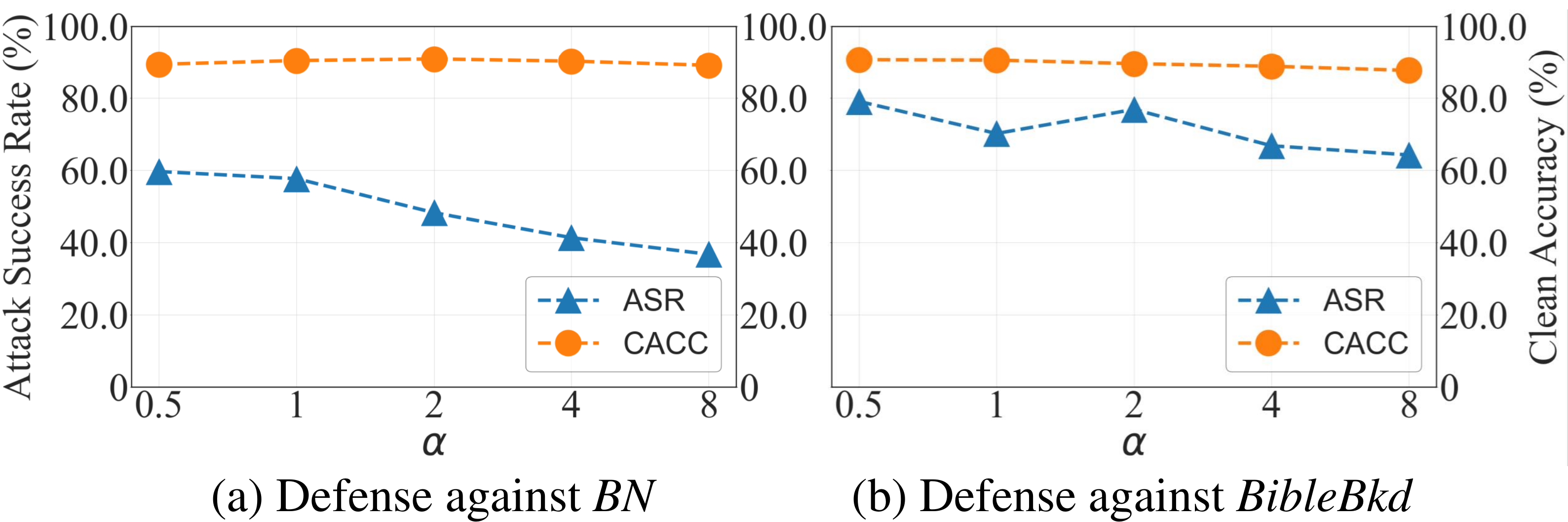}
	\caption{Defense performance of \textbf{NCL} framework against BN and BibleBkd attacks on SST-2 with different $\alpha$ value. 
 }
	\label{fig:alist}
\end{figure}

\noindent\textbf{Hyper-parameter \textbf{$\alpha$} analysis.}
In Eq. \ref{eq:0}, the term $\mathcal{L}_{U C L}$ is the key to mitigating trigger influence. Fig.~\ref{fig:alist} shows the defense result of different $\alpha$ as a weight parameters in Eq. \ref{eq:0}. We find as $\alpha$ increases, the ASR value of the model decreases, while the CACC also decreases. It means that the larger weight of $\mathcal{L}_{U C L}$ term in training, the greater the defense capability of the model against backdoor attacks, meanwhile, the larger the utility decline of the model on benign data.

\subsection{Why Does NCL Work?}\label{sec:loss}

\begin{figure}[htbp]
	\centering
	\includegraphics[width=\linewidth]{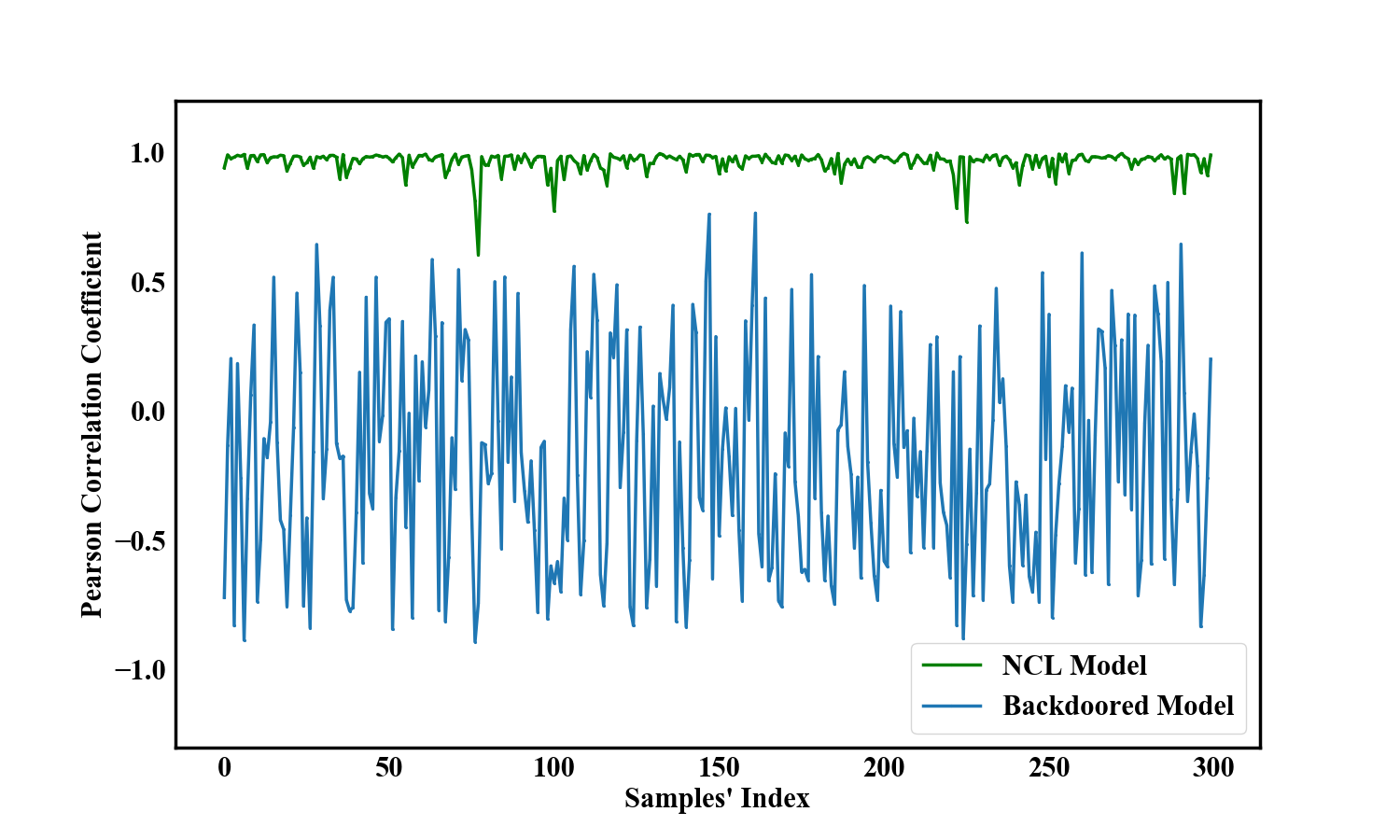}

	\caption{The Pearson Correlation Coefficient of test samples before and after trigger insertion. Green line is the model trained by NCL loss. Blue line is the model trained by cross-entropy loss.
 }
	\label{fig:loss}
\end{figure}

The \textbf{NCL} loss is the most critical part of the framework, which makes the model resistance against backdoor attacks.  We attach additional experiments to prove the effect of \textbf{NCL} loss. In Fig.~\ref{fig:loss}, we trained two BERT models with BN attack using NCL loss without Label-correction (Green line) and cross-entropy loss (Blue line). We obtain sentence embeddings ([cls] output from BERT model) for randomly selected 300 benign examples and their trigger-embedded samples. The we calculated Pearson Correlation Coefficient between benign embeddings and trigger-embedded embeddings for two models. We find that for the model trained with cross-entropy loss, when inserting a trigger word, there is a large swing of sentence embedding. And for the model trained with NCL loss, we observe almost no change. 
It means the model trained with cross-entropy loss, i.e. the backdoored model, is affected by backdoor triggers, while NCL loss avoids this effect.

\section{Conclusion}
In this paper, we propose \textbf{NCL} framework, an effective backdoor defense method, 
which mitigates the impact of backdoor triggers during training utilizing noise-augment method and \textbf{NCL} loss, without the requirement of trigger detection. Experiments show \textbf{NCL} framework achieves outperforming defense performance against mainstream backdoor attacks, while maintaining model utility on benign samples.


\bibliographystyle{IEEEbib}
\bibliography{refs}

\begin{thebibliography}{10}

\bibitem{gu2019badnets}
Tianyu Gu, Kang Liu, Brendan Dolan-Gavitt, and Siddharth Garg,
\newblock ``Badnets: Evaluating backdooring attacks on deep neural networks,''
\newblock {\em IEEE Access}, vol. 7, pp. 47230--47244, 2019.

\bibitem{taneja2022does}
Vardaan Taneja, Pin-Yu Chen, Yuguang Yao, and Sijia Liu,
\newblock ``When does backdoor attack succeed in image reconstruction? a study
  of heuristics vs. bi-level solution,''
\newblock in {\em ICASSP}. IEEE, 2022, pp. 4398--4402.

\bibitem{dai2019backdoor}
Jiazhu Dai, Chuanshuai Chen, and Yufeng Li,
\newblock ``A backdoor attack against lstm-based text classification systems,''
\newblock {\em IEEE Access}, vol. 7, pp. 138872--138878, 2019.

\bibitem{chen2021badnl}
Xiaoyi Chen, Ahmed Salem, Dingfan Chen, Michael Backes, Shiqing Ma, Qingni
  Shen, Zhonghai Wu, and Yang Zhang,
\newblock ``Badnl: Backdoor attacks against nlp models with semantic-preserving
  improvements,''
\newblock in {\em ACSAC}, 2021, pp. 554--569.

\bibitem{qi2021mind}
Fanchao Qi, Yangyi Chen, Xurui Zhang, Mukai Li, Zhiyuan Liu, and Maosong Sun,
\newblock ``Mind the style of text! adversarial and backdoor attacks based on
  text style transfer,''
\newblock in {\em Proceedings of the 2021 Conference on EMNLP}, 2021, pp.
  4569--4580.

\bibitem{qi2021hidden}
Fanchao Qi, Mukai Li, Yangyi Chen, Zhengyan Zhang, Zhiyuan Liu, Yasheng Wang,
  and Maosong Sun,
\newblock ``Hidden killer: Invisible textual backdoor attacks with syntactic
  trigger,''
\newblock in {\em Proceedings of the 59th ACL-IJCNLP}, 2021, pp. 443--453.

\bibitem{xi2021graph}
Zhaohan Xi, Ren Pang, Shouling Ji, and Ting Wang,
\newblock ``Graph backdoor,''
\newblock in {\em 30th USENIX Security Symposium (USENIX Security 21)}, 2021,
  pp. 1523--1540.

\bibitem{kurita2020weight}
Keita Kurita, Paul Michel, and Graham Neubig,
\newblock ``Weight poisoning attacks on pretrained models,''
\newblock in {\em Proceedings of the 58th ACL}, 2020, pp. 2793--2806.

\bibitem{li2021hidden}
Shaofeng Li, Hui Liu, Tian Dong, Benjamin Zi~Hao Zhao, Minhui Xue, Haojin Zhu,
  and Jialiang Lu,
\newblock ``Hidden backdoors in human-centric language models,''
\newblock in {\em Proceedings of ACM CCS}, 2021, pp. 3123--3140.

\bibitem{pan2022hidden}
Xudong Pan, Mi~Zhang, Beina Sheng, Jiaming Zhu, and Min Yang,
\newblock ``Hidden trigger backdoor attack on $\{$NLP$\}$ models via linguistic
  style manipulation,''
\newblock in {\em 31st USENIX Security Symposium (USENIX Security 22)}, 2022,
  pp. 3611--3628.

\bibitem{chen2021mitigating}
Chuanshuai Chen and Jiazhu Dai,
\newblock ``Mitigating backdoor attacks in lstm-based text classification
  systems by backdoor keyword identification,''
\newblock {\em Neurocomputing}, vol. 452, pp. 253--262, 2021.

\bibitem{qi2021onion}
Fanchao Qi, Yangyi Chen, Mukai Li, Yuan Yao, Zhiyuan Liu, and Maosong Sun,
\newblock ``Onion: A simple and effective defense against textual backdoor
  attacks,''
\newblock in {\em Proceedings of ACM CCS}, 2021, pp. 9558--9566.

\bibitem{li2021bfclass}
Zichao Li, Dheeraj Mekala, Chengyu Dong, and Jingbo Shang,
\newblock ``Bfclass: A backdoor-free text classification framework,''
\newblock in {\em Findings of the Association for Computational Linguistics:
  EMNLP}, 2021, pp. 444--453.

\bibitem{krishna2020reformulating}
Kalpesh Krishna, John Wieting, and Mohit Iyyer,
\newblock ``Reformulating unsupervised style transfer as paraphrase
  generation,''
\newblock in {\em Proceedings of the 2020 Conference on EMNLP}, 2020, pp.
  737--762.

\bibitem{gunel2020supervised}
Beliz Gunel, Jingfei Du, Alexis Conneau, and Ves Stoyanov,
\newblock ``Supervised contrastive learning for pre-trained language model
  fine-tuning,''
\newblock {\em arXiv preprint arXiv:2011.01403}, 2020.

\bibitem{socher2013recursive}
Richard Socher, Alex Perelygin, Jean Wu, Jason Chuang, Christopher~D Manning,
  Andrew~Y Ng, and Christopher Potts,
\newblock ``Recursive deep models for semantic compositionality over a
  sentiment treebank,''
\newblock in {\em Proceedings of the 2013 conference on EMNLP}, 2013, pp.
  1631--1642.

\bibitem{zhang2015character}
Xiang Zhang, Junbo Zhao, and Yann LeCun,
\newblock ``Character-level convolutional networks for text classification,''
\newblock {\em Advances in neural information processing systems}, vol. 28,
  2015.

\bibitem{devlin2019bert}
Jacob Devlin, Ming-Wei Chang, Kenton Lee, and Kristina Toutanova,
\newblock ``Bert: Pre-training of deep bidirectional transformers for language
  understanding,''
\newblock in {\em NAACL-HLT}, 2019, pp. 4171--4186.

\bibitem{liu2019roberta}
Yinhan Liu, Myle Ott, Naman Goyal, Jingfei Du, Mandar Joshi, Danqi Chen, Omer
  Levy, Mike Lewis, Luke Zettlemoyer, and Veselin Stoyanov,
\newblock ``Roberta: A robustly optimized bert pretraining approach,''
\newblock {\em arXiv preprint arXiv:1907.11692}, 2019.

\bibitem{sanh2019distilbert}
Victor Sanh, Lysandre Debut, Julien Chaumond, and Thomas Wolf,
\newblock ``Distilbert, a distilled version of bert: smaller, faster, cheaper
  and lighter,''
\newblock {\em arXiv preprint arXiv:1910.01108}, 2019.

\bibitem{gao2021simcse}
Tianyu Gao, Xingcheng Yao, and Danqi Chen,
\newblock ``Simcse: Simple contrastive learning of sentence embeddings,''
\newblock in {\em Proceedings of the 2021 Conference on EMNLP}, 2021, pp.
  6894--6910.

\bibitem{iyyer2018adversarial}
Mohit Iyyer, John Wieting, Kevin Gimpel, and Luke Zettlemoyer,
\newblock ``Adversarial example generation with syntactically controlled
  paraphrase networks,''
\newblock in {\em NAACL-HLT}, 2018, pp. 1875--1885.

\end{thebibliography}
\end{document}